\documentclass[12pt]{article}
\usepackage{graphicx}
\usepackage{amsmath}
\usepackage{amssymb}
\usepackage{caption2}
\begin{document}
\bibliographystyle{prsty}
\begin{center}
{\large {\bf \sc{  Analysis of  Y(2175) as a tetraquark state with QCD sum rules }}} \\[2mm]
Zhi-Gang Wang \footnote{E-mail,wangzgyiti@yahoo.com.cn.  }     \\
 Department of Physics, North China Electric Power University,
Baoding 071003, P. R. China

\end{center}

\begin{abstract}
In this article, we take the point of view that $Y(2175)$ be a
tetraquark state which consists of   color octet constituents, and
calculate its mass and decay constant within the framework of the
QCD sum rule approach.  We release  standard criterion
  in the QCD sum rules approach and take more phenomenological analysis,  the  value of
  the mass of   $Y(2175)$ is consistent with experimental data; there may be
some tetraquark  components in  the  state $Y(2175)$. If we retain
 standard criterion, larger mass than  experimental data can
be obtained, the current $J_\mu(x)$ can interpolate a tetraquark
state with larger mass, or  $Y(2175)$ has some components with
larger mass. The dominating contribution comes from the perturbative
term, which is in contrast to the sum rules with  interpolating
currents constructed from diquark pairs.   The tetraquark states may
consist of color octet constituents rather than  diquark pairs.
\end{abstract}



\section{Introduction}

Recently,  Babar collaboration observed a resonance with the quantum
numbers $J^{PC}=1^{--}$ near the threshold in  process $e^+ e^- \to
\phi f_0(980)$ via initial-state radiation \cite{BABAR}.
Breit-Wigner mass is $m_Y = (2.175 \pm 0.010\pm 0.015) GeV$ and
width is narrow $\Gamma_Y = (58\pm 16\pm 20 )MeV$.  The resonance
may be interpreted as an $\bar{s}s$ analogue of  $Y(4260)$, or as an
$\bar{s}s\bar{s}s$ state that decays predominantly to  $\phi
f_0(980)$. In this article, we take the point of view that the state
$Y(2175)$ (thereafter we take the notation $Y(2175)$) be a
tetraquark state with the quantum numbers $J^{PC}=1^{--}$,  and
calculate its mass and decay constant in the framework of the QCD
sum rules approach \cite{SVZ79,Narison89,QCDreview}. In the QCD sum
rules, operator product expansion is used to expand the time-ordered
currents into a series of quark and gluon condensates  which
parameterize  long distance properties of  the QCD vacuum. Based on
current-hadron duality, we can obtain copious information about the
hadronic parameters at phenomenological side.

Whether or not there exists  a tetraquark configuration
$\bar{s}s\bar{s}s$ which can result in  baryonium state  is of great
importance   itself, because it provides a new opportunity for a
deeper understanding of  low energy QCD. We explore this
possibility, later experimental data can confirm or reject this
assumption.   Interactions of  one-gluon exchange  and direct
instantons lead to significant attractions between the quarks in
$0^+$ channel,
  $Y(4260)$ can be taken as consist of  scalar diquark ($\epsilon_{kij}c^T_i C
\gamma_5 s_j$) pairs in relative $P$-wave \cite{Maiani05}. However,
two $s$ quarks cannot cluster together to form a scalar diquark, if
$Y(2175)$ is the cousin of  $Y(4260)$, why they have so different
substructures?

 Existence of the tetraquark states has not been confirmed with experimental data yet, however, there are
 evidences for those  exotic states.   Numerous
candidates  with the same quantum numbers $J^{PC}=0^{++}$ below $2
GeV$ can not be accommodated in one $q\bar{q}$ nonet,  some are
supposed to be glueballs, molecules and multiquark states
\cite{Close2002,ReviewScalar}. There maybe different dynamics that
dominate   $0^{++}$ mesons  below and above $1GeV$, which results in
two scalar nonets below $1.7 GeV$. Attractive interactions of
one-gluon exchange favor  formation of diquarks in  color
antitriplet $\overline{3}_{ c}$, flavor antitriplet $\overline{3}_{
f}$ and spin singlet  $ 1_{s} $. Strong attractions between the
states $(qq)_{\overline{3}}$ and $(\bar{q}\bar{q})_{3}$ in $S$-wave
may result in a nonet manifested below $1GeV$,  while the
conventional $^3P_0$ $\bar{q}q$ nonet would have masses about
$1.2-1.6 GeV$. Furthermore, there are enough candidates for $^3P_0$
$\bar{q}q$ nonet mesons, $a_0(1450)$, $f_0(1370)$, $K^*(1430)$,
$f_0(1500)$ and $f_0(1710)$ \cite{Close2002,ReviewScalar}. If we
take scalar diquarks  $U^a= \epsilon_{abc} d_b^T(x)C\gamma_5
s_c(x)$, $D^a=\epsilon_{abc} u_b^T(x)C\gamma_5 s_c(x)$ and
$S^a=\epsilon_{abc} u_b^T(x)C\gamma_5 d_c(x)$ as  basic
constituents, the mass formula of $0^{++}$ nonet mesons below $1GeV$
can be naturally  explained. Comparing with the traditional ${\bar
q} q $ nonet mesons, the mass spectrum is inverted.  The lightest
state is the non-strange isosinglet ($\bar {S}^a S^a$), the heaviest
are the degenerate isosinglet and isovectors with hidden $\bar s s$
pairs, while the four strange states lie in between
\cite{Close2002,ReviewScalar}. The mass spectrum of the scalar nonet
mesons as tetraquark states below $1GeV$ has been studied  with the
QCD sum rules approach \cite{Brito05,Wang05}, for more literature on
the tetraquark states consist of diquark pairs with the QCD sum
rules approach, one can consult e.g. Refs.\cite{TetraquarkS,Wang06}.

The article is arranged as follows:  we derive the QCD sum rules for
the mass and decay constant of   $Y(2175)$  in section 2; in section
3, numerical results and discussions; section 4 is reserved for
conclusion.

\section{QCD sum rules for the mass of the $Y(2175)$}
In the following, we write down  the two-point correlation function
$\Pi_{\mu\nu}(p^2)$ in the QCD sum rules approach,
\begin{eqnarray}
\Pi_{\mu\nu}(p^2)&=&i\int d^4x e^{ip \cdot x} \langle
0|T\{J_\mu(x)J_\nu^+(0)\}|0\rangle \, ,  \\
J_\mu(x)&=&\bar{s}(x)\gamma_\mu \lambda^a s(x) \bar{s}(x) \lambda^a s(x)\, ,\\
f_Ym^4_Y\epsilon_\mu&=&\langle 0|J_\mu(0)|Y\rangle \, .
\end{eqnarray}
Where  $\lambda^a$'s are  Gell-Mann matrixes for color $SU(3)$
group,  $\epsilon_\mu$ and  $f_Y$ stand for the polarization vector
and decay constant of  $Y(2175)$, respectively.  We take  color
octet operators $\bar{s}(x)\gamma_\mu \lambda^a s(x) $ and $
\bar{s}(x) \lambda^a s(x)$ as   basic constituents in constructing
the vector current $J_{\mu}(x)$. Originally, color octet operators
$\bar{s}\lambda^a s$, $\bar{q}\lambda^a q$ ,
$\bar{s}\gamma_5\lambda^a s$ and $\bar{q}\gamma_5\lambda^a q$ were
used to construct the interpolating currents for the scalar mesons
$a_0(980)$ and $f_0(980)$ as tetraquark states \cite{Latorre85}.
There are other two vector current operators $J^A_\mu(x)$ and
$J^B_\mu(x)$ with the same quantum numbers $J^{PC}=1^{--}$ as
$Y(2175)$,
\begin{eqnarray}
J^A_\mu(x)&=&\bar{s}(x)\gamma_\mu   s(x) \bar{s}(x)   s(x) \, ,
\nonumber\\
 J^B_\mu(x)&=&\epsilon^{kij}\epsilon^{kmn}\left\{s^T_i(x) C
\sigma_{\mu\nu}s_j(x)\bar{s}_m(x)C\gamma^\nu
\gamma_5\bar{s}^T_n(x)\right.\nonumber\\
&&\left.+\bar{s}_i(x) C
\sigma_{\mu\nu}\bar{s}^T_j(x)s^T_m(x)C\gamma^\nu
\gamma_5s_n(x)\right\} \, .
\end{eqnarray}
Where    $k$, $i$, $j$, $m$, $n$ are  color indexes,  $C$ is charge
conjunction matrix,  $\mu$ and $\nu$ are   Lorentz indexes.   If we
take  color singlet operators $\bar{s}(x)\gamma_\mu s(x) $ and $
\bar{s}(x)
  s(x)$ as  basic constituents, and choose the current operator
$J^A_\mu(x)$, which can interpolate a tetraquark state, whether
compact state or  loose deuteron-like $\phi f_0(980)$ bound state,
it is difficult to separate the contributions of  bound state from
 scattering $\phi f_0(980)$ state. In this article, we take
$Y(2175)$ as a baryonium state and choose the current $J_\mu(x)$,
although  $J^A_\mu(x)$  has non-vanishing coupling with $Y(2175)$.
In the diquark-antidiquark model \cite{Maiani05},  $Y(4260)$ is
taken as consist of  scalar diquark ($\epsilon_{kij}c^T_i C \gamma_5
s_j$) pairs in relative $P$-wave. One can take  $Y(2175)$ as  the
cousin of $Y(4260)$, decays  $Y(4260) \to J/\psi f_0(980)$ and
$Y(2175) \to \phi f_0(980)$ occur with the same mechanism, however,
 $Y(2175)$ can not be constructed from  scalar $ss$ diquark
pairs, because two $s$ quarks can not cluster together to form a
scalar diquark due to  Fermi statistics, we have to resort to the
constituents, a tensor diquark and a vector diquark in relative
$S$-wave, to construct the interpolating current, if one insist on
that the multiquark current operators should be constructed from
diquark pairs \cite{Brito05,TetraquarkS}. It is odd that the cousins
have very different substructures,  $Y(4260)$ may have structure
$\bar{c}\gamma_\mu \lambda^ac \bar{s} \lambda^a s +\bar{s}\gamma_\mu
\lambda^as \bar{c} \lambda^a c $.

The correlation function  $\Pi_{\mu\nu}(p)$ can be decomposed as
follows:
\begin{eqnarray}
\Pi_{\mu\nu}(p)&=&-\Pi_1(p^2)\left(g_{\mu\nu}-\frac{p_\mu
p_\nu}{p^2}\right)+\Pi_0(p^2)\frac{p_\mu p_\nu}{p^2},
\end{eqnarray}
due to  Lorentz covariance. The invariant functions $\Pi_1$ and
$\Pi_0$ stand for the contributions from the vector and scalar
mesons, respectively. In this article, we choose the tensor
structure $g_{\mu\nu}-\frac{p_\mu p_\nu}{p^2}$ to study the mass of
the vector meson.

 According to   basic assumption of current-hadron duality in
the QCD sum rules approach \cite{SVZ79}, we insert  a complete
series of intermediate states satisfying  unitarity   principle with
the same quantum numbers as the current operator $J_\mu(x)$
 into the correlation function in
Eq.(1)  to obtain the hadronic representation. After isolating the
pole term  of the lowest state $Y(2175)$, we obtain the following
result:
\begin{eqnarray}
\Pi_{\mu\nu}(p^2)&=&i\int d^4x e^{ip \cdot x} \langle
0|J_\mu(x) \sum_{n,\epsilon_n} \int  \frac{d^3q}{(2\pi)^3 2E_n} |q_n,\epsilon_n\rangle\langle q_n,\epsilon_n
| J_\nu^+(0)|0\rangle \Theta(t)\, , \nonumber \\
 &=&-\frac{f_Y^2m_Y^8}{m_Y^2-p^2}\left\{g_{\mu\nu}-\frac{p_\mu
p_\nu}{p^2}\right\}+\cdots \, ,\\
\frac{\mbox{Im}\Pi_1(s)}{\pi}&=&f_Y^2m_Y^8+\rho_{QCD}\Theta(s-s_0)
\, .
\end{eqnarray}
The intermediate states are saturated by the states
$|q_n,\epsilon_n\rangle$ with the same quantum numbers as $Y(2175)$,
 high resonances (if there are some) $Y_1$, $Y_2$, $Y_3$ $\cdots$
appear consequentially before   continuum states (which can be
described by the  contributions from asymptotic quarks  and gluons)
set on. If we choose standard criterion that the dominating
contribution comes from the pole terms, $Y_1$, $Y_2$, $\cdots$
should be concluded in besides $Y(2175)$. It is difficult to analyze
those terms qualitatively or quantitatively without experimental
data.  We approximate the hadronic spectral density $\Pi_1(s)$ with
the corresponding one $\rho_{QCD}$ from  perturbative QCD above the
threshold $s_0$. We will revisit this subject in next section.

In the following, we briefly outline  operator product expansion for
the correlation function $\Pi_{\mu\nu }(p)$  in perturbative QCD
theory. The calculations are performed at   large space-like
momentum region $p^2\ll 0$, which corresponds to small distance
$x\approx 0$ required by   validity of   operator product expansion
approach. We write down the "full" propagator $S_{ab}(x)$ of a
massive light quark in the presence of the vacuum condensates
firstly \cite{SVZ79}\footnote{One can consult the last  article of
Ref.\cite{SVZ79} for technical details in deriving the full
propagator.},
\begin{eqnarray}
S_{ab}(x)&=& \frac{i\delta_{ab}\!\not\!{x}}{ 2\pi^2x^4}
-\frac{\delta_{ab}m_s}{4\pi^2x^2}-\frac{\delta_{ab}}{12}\langle
\bar{s}s\rangle +\frac{i\delta_{ab}}{48}m_s
\langle\bar{s}s\rangle\!\not\!{x}-\nonumber\\
&&\frac{\delta_{ab}x^2}{192}\langle \bar{s}g_s\sigma Gs\rangle
 +\frac{i\delta_{ab}x^2}{1152}m_s\langle \bar{s}g_s\sigma
 Gs\rangle \!\not\!{x}-\nonumber\\
&&\frac{i}{32\pi^2x^2}\frac{\lambda^A_{ab}}{2} G^A_{\mu\nu}
(\!\not\!{x} \sigma^{\mu\nu}+\sigma^{\mu\nu} \!\not\!{x})  +\cdots
\, ,
\end{eqnarray}
where $\langle \bar{s}g_s\sigma Gs\rangle=\langle
\bar{s}g_s\sigma_{\alpha\beta} G^{\alpha\beta}s\rangle$ , then
contract the quark fields in the correlation function
$\Pi_{\mu\nu}(p)$ with Wick theorem, and obtain the result:
\begin{eqnarray}
\Pi_{\mu\nu}(p)&=&i \lambda^a_{ij}\lambda^a_{mn}
\lambda^b_{i'j'}\lambda^b_{m'n'} \int d^4x \, e^{i p
\cdot x} \nonumber\\
&&\left\{Tr\left[ \gamma_\mu S_{ji'}(x)\gamma_\nu S_{j'i}(-x)\right]Tr\left[   S_{nm'}(x)  S_{n'm}(-x)\right] \right.\nonumber \\
&&\left. +Tr\left[\gamma_\mu S_{jm'}(x)S_{n'i}(-x)\right]Tr\left[
S_{ni'}(x)\gamma_\nu S_{j'm}(-x)\right] \right\}\, .
\end{eqnarray}
Substitute the full $s$ quark propagator into above correlation
function and complete  integral in  coordinate space, we can obtain
the correlation function $\Pi_1$ at the level of quark-gluon degree
of freedom:

\begin{eqnarray}
\Pi_1&=& -\frac{p^8}{27648\pi^6}\mbox{log}(-p^2)-\frac{m_s\langle
\bar{s}s\rangle p^4}{72\pi^4}\mbox{log}(-p^2)
+\frac{p^4}{13824\pi^4} \langle \frac{\alpha_s GG}{\pi}\rangle
\mbox{log}(-p^2) \nonumber \\
&&-\frac{\langle\bar{s}s\rangle^2
p^2}{27\pi^2}\mbox{log}(-p^2)+\frac{m_s\langle\bar{s}g_s \sigma
 Gs\rangle
p^2}{216\pi^4}\mbox{log}(-p^2)+\frac{\langle\bar{s}s\rangle \langle
\bar{s}g_s \sigma  G s\rangle}{18\pi^2}\mbox{log}(-p^2) \nonumber\\
&&-\frac{40m_s\langle\bar{s}s\rangle^3}{27p^2} -\frac{\langle
\bar{s}g_s \sigma G s \rangle^2}{72 \pi^2 p^2} -\frac{16m_s\langle
\bar{s}s\rangle^2 \langle \bar{s}g_s \sigma  G s\rangle}{81p^4}
+\cdots \, ,
\end{eqnarray}
 where $\langle \frac{\alpha_s GG}{\pi}\rangle=\langle \frac{\alpha_s
 G_{\alpha\beta}G^{\alpha\beta}}{\pi}\rangle$. We carry out  operator
product expansion to the vacuum condensates adding up to
dimension-11. In calculation, we
 take  assumption of vacuum saturation for  high
dimension vacuum condensates, they  are always
 factorized to lower condensates with vacuum saturation in the QCD sum rules,
  factorization works well in  large $N_c$ limit.
In this article, we take into account the contributions from the
quark condensate $\langle \bar{s}s \rangle$,  mixed condensate
$\langle \bar{s}g_s \sigma  G{s} \rangle $, gluon condensates
$\langle \frac{\alpha_s GG}{\pi}\rangle$, and neglect the
contributions  from other high dimension condensates (for example,
$\langle g_s^3 G^3\rangle$), which are suppressed by large
denominators and would not play significant roles.

Once  analytical results are obtained,
  then we can take  current-hadron duality  below the threshold
$s_0$ and perform  Borel transformation with respect to  variable
$P^2=-p^2$, finally we obtain  the following sum rule:

\begin{eqnarray}
 f_Y^2 m_Y^8e^{-\frac{m_Y^2}{M^2}}&=&\int_{16m_s^2}^{s_0}dt
e^{-\frac{t}{M^2}}\frac{\mbox{Im}\Pi(t)}{\pi}
+\frac{\langle\bar{s}g_s \sigma G s\rangle^2}{72\pi^2
}+\frac{40m_s\langle \bar{s}s\rangle^3}{27 }-\frac{16
m_s\langle\bar{s}s\rangle^2 \langle\bar{s}g_s \sigma G s\rangle}{81
M^2} \, , \nonumber \\
\end{eqnarray}

\begin{eqnarray}
\frac{\mbox{Im}\Pi(t)}{\pi}&=& \frac{t^4}{27648\pi^6}
+\frac{m_s\langle \bar{s}s\rangle t^2}{72 \pi^4}-\frac{t^2}{13824
\pi^4}\langle \frac{\alpha_s GG}{\pi} \rangle -\frac{m_s
\langle\bar{s}g_s \sigma G s\rangle t}{216\pi^4}
 \nonumber \\
&& +\frac{\langle\bar{s}s\rangle^2t}{27\pi^2}-\frac{\langle\bar{s}
s\rangle\langle\bar{s}g_s\sigma G s\rangle }{18 \pi^2} \, .
\end{eqnarray}

Differentiate the above sum rule with respect to   variable
$\frac{1}{M^2}$, then eliminate the quantity $f_Y$, we obtain the
QCD sum rule for the mass:
\begin{eqnarray}
  m_Y^2&=& \left\{\int_{16m_s^2}^{s_0}dt t
e^{-\frac{t}{M^2}}\frac{\mbox{Im}\Pi(t)}{\pi}
  +\frac{16
m_s\langle\bar{s}s\rangle^2 \langle\bar{s}g_s \sigma G s\rangle}{81
} \right\}/ \nonumber\\
&&\left\{\int_{16m_s^2}^{s_0}dt
e^{-\frac{t}{M^2}}\frac{\mbox{Im}\Pi(t)}{\pi}
+\frac{\langle\bar{s}g_s \sigma G s\rangle^2}{72\pi^2
}+\frac{40m_s\langle ss\rangle^3}{27 } -\frac{16
m_s\langle\bar{s}s\rangle^2 \langle\bar{s}g_s \sigma G s\rangle}{81
M^2} \right\} \, . \nonumber \\
\end{eqnarray}

It is easy to integrate over  the variable  $t$, we prefer this
formulation for simplicity. From Eq.(13), we can obtain the mass
$m_Y$, then take $m_Y$ as input parameter, we   obtain the decay
constant $f_Y$ from Eq.(11) with the same values of the vacuum
condensates.
\section{Numerical results and discussions}
The input parameters are taken to be the standard values $\langle
\bar{q}q \rangle=-(0.24\pm 0.02 GeV)^3$, $\langle \bar{s}s
\rangle=(0.8\pm 0.2 )\langle \bar{q}q \rangle$, $\langle
\bar{s}g_s\sigma Gs \rangle=m_0^2\langle \bar{s}s \rangle$,
$m_0^2=(0.8 \pm 0.2)GeV^2$, $\langle \frac{\alpha_s
GG}{\pi}\rangle=(0.33GeV)^4 $
 and $m_s=(0.14\pm0.02)GeV$ \cite{SVZ79,Narison89,QCDreview}.
 For the multiquark  states,  the
contribution from  terms with the gluon condensate $\langle
\frac{\alpha_s GG}{\pi}\rangle $ is of minor importance
\cite{Wang05,Wang06,Oka94}. The contribution from  $\langle
\frac{\alpha_s GG}{\pi}\rangle $ in Eq.(11) is less than $2\%$, and
uncertainty is neglected here.

The main contribution in Eq.(11) comes from the perturbative term,
 (a piece of)  standard criterion of the QCD sum rules can be
satisfied; which is in contrast to the ordinary sum rules with the
interpolating currents constructed from the multiquark
configurations, where the contribution comes from the perturbative
term is very small \cite{Narison04}, the main contributions come
from the terms with the quark condensates $\langle \bar{q}q \rangle$
and $\langle \bar{s}s\rangle$, sometimes the mixed condensates
$\langle \bar{q}g_s\sigma G q \rangle$ and $\langle \bar{s}g_s\sigma
G s\rangle$ also play important roles \cite{Wang05,Wang06,Oka94}.

The values of the vacuum condensates have been updated with
experimental data for  $\tau$ decays, the QCD sum rules for the
baryon masses and analysis of the charmonium spectrum
\cite{Zyablyuk, Ioffe2005,AlphaS}. As the main contribution comes
from the perturbative term, uncertainties of the vacuum condensates
can only result in very small uncertainty for numerical values of
the mass $m_Y$ and decay constant $f_Y$, the standard values and
updated values of the vacuum condensates can only lead to results of
minor difference, we choose the standard values of the vacuum
condensates in the calculation.

Neglecting the contributions from the vacuum condensates and taking
the parameters $\sqrt{s_0}=2.4GeV$, $M^2=(3-7)GeV^2$, we can obtain
the value $m_Y=2.17GeV$. It is indeed the main contribution comes
from the perturbative term. If we take  color octet operators
$\bar{q}\lambda^a q$, $\bar{q}i\gamma_5\lambda^a q$,
$\bar{q}\gamma_\mu\lambda^a q$, $\bar{q}\gamma_\mu \gamma_5\lambda^a
q$ and $\bar{q}\sigma_{\mu\nu}\lambda^a q$ as  basic constituents to
construct the tetraquark currents, the contributions of the
perturbative terms may have   dominant contributions, in other
words, the tetraquark states may consist of  color octet
constituents rather than   diquark pairs
\cite{Maiani05,Brito05,Wang05,TetraquarkS,Wang06}.

For  the conventional (two-quark) mesons and (three-quark) baryons,
the hadronic  spectral densities are experimentally well known,
separation between the ground state and excited states is large
enough, the "single-pole + continuum states" model works well in
representing the phenomenological spectral densities. The continuum
states can be approximated by the contributions from the asymptotic
quarks and gluons, and the single-pole dominance condition can be
well satisfied,
\begin{eqnarray}
\int_{s_0}^{\infty}\rho_{A}e^{-\frac{s}{M^2}}ds <
\int^{s_0}_{0}(\rho_{A}+\rho_{B})e^{-\frac{s}{M^2}}ds \, ,
\end{eqnarray}
 where  $\rho_{A}$ and
$\rho_{B}$ stand for the contributions from the perturbative and
non-perturbative part of the spectral density, respectively. From
  criterion  in Eq.(14), we can obtain the maximal value of the Borel parameter
$M_{max}$, exceed this value,  single-pole dominance will be
spoiled. On the other hand, the Borel parameter must be chosen large
enough to warrant  convergence of  operator product expansion and
the contributions from the high dimension vacuum condensates, which
are  known poorly,  are of minor importance, the minimal value of
the Borel parameter $M_{min}$ can be determined.

For the conventional  mesons and  baryons, the Borel window
$M_{max}-M_{min}$  is rather large and  reliable QCD sum rules can
be obtained. However, for the multiquark states i.e. tetraquark
states, pentaquark states, hexaquark states, etc, the spectral
densities $\rho\sim s^n$ with $n$ is larger than the
 ones for the conventional hadrons,     integral
$\int_0^{\infty} s^n \exp \{-\frac{s}{M^2}\} ds$ converges more
slowly \cite{Narison04}. If one do not want to release the criterion
in Eq.(14), we have to either postpone the threshold parameter $s_0$
to very large value or choose very small value for the Borel
parameter $M_{max}$. With large value for the threshold parameter
$s_0$ , for example, $s_0 \gg M_{gr}^2$, here  $gr$ stands for the
ground state, the contributions from  high resonance states and
continuum states are included in, we cannot use  single-pole (or
ground state) approximation for the spectral densities; on the other
hand, with very small value for the Borel parameter $M_{max}$,
operator product expansion is broken down, and the Borel window
$M_{max}-M_{min}$ shrinks to zero or negative values. We should
resort to  "multi-pole $+$ continuum states" to approximate the
phenomenological spectral densities. Onset of the continuum states
is not abrupt,   the ground state, the first excited state, the
second excited state, etc, the continuum states appear sequentially;
the excited states may be loose  bound states and have large widths.
The threshold parameter $s_0$  is postponed to large value, at that
energy scale, the spectral densities can be well approximated by the
contributions from the asymptotic quarks and gluons, and of minor
importance for the sum rules\cite{Wang06}.

From Figs.1-2, we can see that the main contribution comes from the
perturbative term, the hadronic spectral density above and below the
threshold can be successfully approximated by the perturbative term.
If we take typical values for the parameters $\sqrt{s_0}=2.4GeV$ and
$M^2=4.0GeV^2$, the contributions from  continuum states  are
dominating,
\begin{eqnarray}
\frac{\int_0^{s_0}dt t^4 e^{-t/M^2}}{\int_0^{\infty}dt t^4
e^{-t/M^2}} < 2\% \, .
\end{eqnarray}
It  is  not an indication that  non-existence  of the tetraquark
states due to lack experimental information about physics above the
threshold $s_0$. One may  refuse   the value extracted from
continuum dominating QCD sum rules as quantitatively reliable if one
insists on that contribution from the pole term should be larger
than (or about) $50\%$ in conventional QCD sum rules.

In this article, we cannot  find the conventional Borel window (or
the Borel window is too small to make robust prediction) and
threshold parameter for the sum rule in Eq.(11); and release
standard criterion and prefer more phenomenological analysis. We
choose the suitable values for the Borel parameter $M$, on the one
hand, the minimal values $M_{min}$ are large enough to warrant the
convergence of  operator product expansion, for
$M_{min}>\sqrt{3}GeV$, the dominating contribution comes from the
perturbative term in Eq.(11), larger than $90\%$; on the other hand,
the maximal values $M_{max}$ are small enough to suppress the
contributions from the high resonance (or excited) states and
continuum states,
 we choose naive analysis  $e^{-s_0/(M_{max})^2}\leq e^{-1}$.

\begin{figure}
 \centering
 \includegraphics[totalheight=7cm,width=13cm]{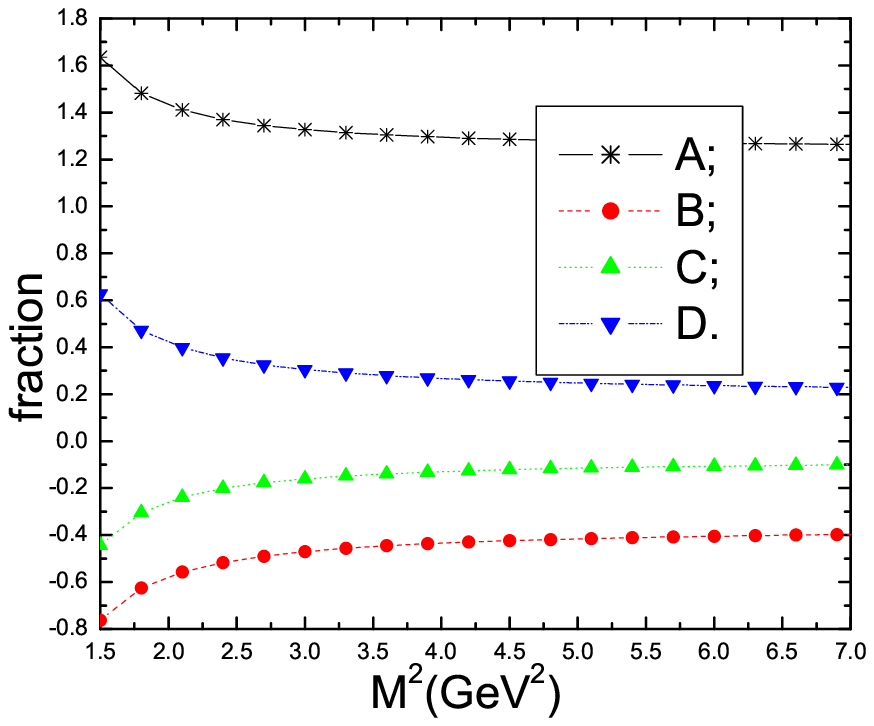}
  \caption{Fractions of different terms with  Borel parameter $M^2$ in the right hand side of
  Eq.(11).
  $A$, $B$, $C$ and $D$ stand for the contributions from the  terms proportional to $t^4$, $\langle \bar{s}s\rangle$,
  $\langle \bar{s}s\rangle\langle \bar{s}g_s \sigma Gs\rangle$ and $\langle \bar{s}s\rangle^2$, respectively. The contributions from
  the terms of minor importance (less than $10\%$) are not shown explicitly. The   input parameters are
  taken as $\sqrt{s_0}=2.4GeV$, and  central values of the vacuum condensates. }
\end{figure}

\begin{figure}
 \centering
 \includegraphics[totalheight=7cm,width=13cm]{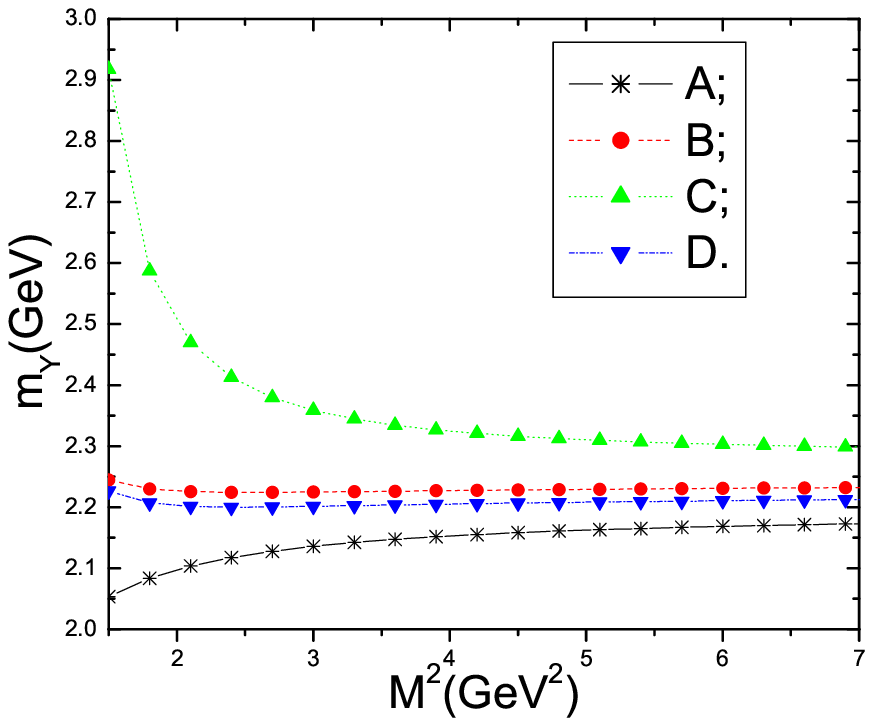}
  \caption{Mass from different terms with  Borel parameter $M^2$ in Eq.(13),
  $A$, $B$, $C$ and $D$ from the  terms proportional to $t^4$, $t^4+\langle \bar{s}s\rangle$,
  $t^4+\langle \bar{s}s\rangle+\langle \bar{s}s\rangle\langle \bar{s}g_s \sigma Gs\rangle$ and
  $t^4+\langle \bar{s}s\rangle+\langle \bar{s}s\rangle\langle \bar{s}g_s \sigma Gs\rangle+\langle \bar{s}s\rangle^2$, respectively. The contributions from
  the terms of minor importance are not shown explicitly. The   input parameters are
  taken as $\sqrt{s_0}=2.4GeV$, and  central values of the vacuum condensates.}
\end{figure}

In Figs.3-4, we plot the values of the mass $m_Y$ and decay constant
$f_Y$ with variation of the threshold parameter $s_0$. From those
figures, we can see that the values increase steadily with increase
of $s_0$, the QCD sum rules cannot indicate existence of the
tetraquark state $Y(2175)$ strictly, we should adopt more
phenomenological analysis.

The vector current $J^c_\mu=\bar{c}\gamma_\mu c$ can interpolate the
vector mesons $J/\psi(1S)$, $\psi(2S)$, $\psi(3770)$, $\psi(4040)$,
$\psi(4160)$ and $\psi(4415)$, the  correlation function
$\Pi^c_{\mu\nu}$,
\begin{eqnarray}
\Pi^c_{\mu\nu}=i\int d^4x e^{ip\cdot x} \langle 0|T\left\{J^c_\mu(x)
J^c_\nu(0)\right\}|0\rangle \, ,
\end{eqnarray}
 can be saturated by
$J/\psi(1S)$, $\psi(2S)$, $\psi(3770)$, $\psi(4040)$, $\psi(4160)$,
$\psi(4415)$ and continuum states at the phenomenological side
\cite{IoffeJ}. The masses and widths of those vector mesons are well
known, one can consult  PDG for details \cite{PDG}. If experimental
data about the higher resonances $Y_1$, $Y_2$, $\cdots$ are
available (suppose there are some), we can make analogous analysis
as in the vector hidden charm channels to avoid difficulty in
choosing the Borel parameter $M^2$ and  threshold parameter $s_0$.

 However,  present experimental knowledge about the
phenomenological hadronic spectral densities of the multiquark
states is  rather vague, even existence of the multiquark states is
not confirmed with confidence, and no knowledge about either there
are high resonances or not. Criterion in Eq.(14) cannot lead to
reasonable Borel parameter $M$ and threshold parameter $s_0$ for the
multiquark states, we can either reject the QCD sum rules for the
multiquark states or release the condition, we are optimistical
participators\footnote{We take the point of view that although the
standard criterion of the QCD sum rules cannot be satisfied for the
multiquark states, experimental data about the high resonances is of
great importance; we should  analyze the ground state and high
resonances together and  come out the difficulty. }.

 In this article, we approximate the spectral density with
the contribution from the single-pole term, the threshold parameter
$s_0$  is taken slightly above the ground state mass (
$\sqrt{s_0}>m_{Y}+\frac{\Gamma_{Y}}{2}$) to subtract the
contributions from the high resonances   and continuum states. We
take $\sqrt{s_0}= (2.3-2.5)GeV
>2.2GeV$, it is reasonable for  Breit-Wigner mass  $m_Y = 2.175
\pm 0.010\pm 0.015 GeV$ and  width $\Gamma_Y = 58\pm 16\pm 20 MeV$.
The Borel parameter $M$ can be chosen to be $M^2=(3.0-7.0)GeV^2$, in
this region, the values of the mass and decay constant are rather
stable with respect to variation of the Borel parameter, which are
shown in Fig.3 and Fig.4 respectively.
\begin{figure}
 \centering
 \includegraphics[totalheight=7cm,width=13cm]{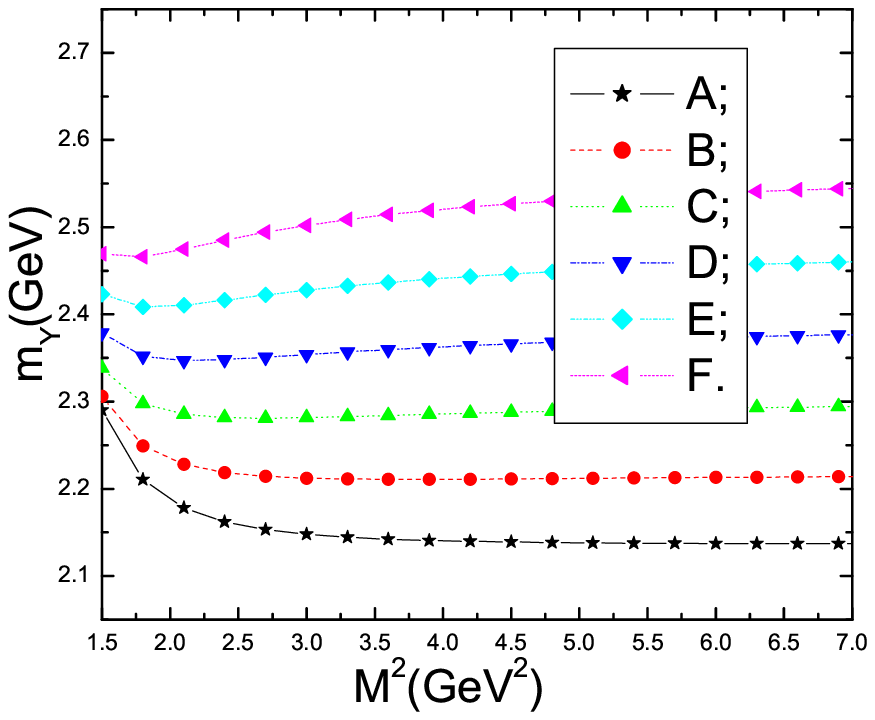}
  \caption{  $m_Y$  with  the central values of the vacuum condensates. $A$, $B$, $C$, $D$, $E$ and $F$ stand for
  $\sqrt{s_0}=$ $2.3GeV$, $2.4GeV$, $2.5GeV$, $2.6GeV$, $2.7 GeV$ and  $2.8GeV$ respectively.}
\end{figure}
\begin{figure}
 \centering
 \includegraphics[totalheight=7cm,width=13cm]{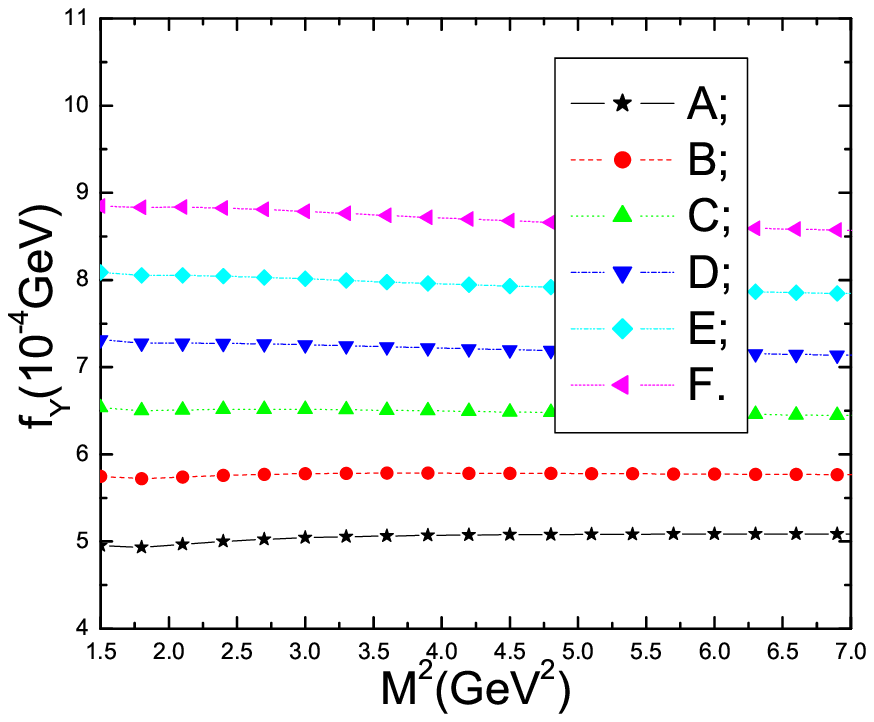}
  \caption{  $f_Y$  with  the central values of the vacuum  condensates. $A$, $B$, $C$, $D$, $E$ and $F$ stand for
  $\sqrt{s_0}=$ $2.3GeV$, $2.4GeV$, $2.5GeV$, $2.6GeV$, $2.7 GeV$ and  $2.8GeV$ respectively.}
\end{figure}
Finally, we obtain the values of the mass and  decay constant  of
 $Y(2175)$,
\begin{eqnarray}
m_Y&=&(2.21\pm0.09)GeV \, , \nonumber\\
f_Y&=&(5.78\pm 0.89)10^{-4}GeV \, .
\end{eqnarray}
The main uncertainty   comes from the threshold parameter $s_0$,
uncertainties of the vacuum condensates and $m_s$ can only lead to
minor uncertainty.

If we take smaller values for the Borel parameter $M^2$ and larger
values for the threshold parameter $s_0$, for example,
$M^2=(1.5-3.0)GeV^2$ and $\sqrt{s_0}=(2.3-2.9)GeV$, the contribution
from the pole term (or ground state) can be greatly enhanced, which
is shown in table.1. From the table, we can see that if we take
$M^2=(1.5-2.0)GeV^2$  and $\sqrt{s_0}=(2.6-2.8)GeV$ , the
contribution from the pole term (or ground state) is about
$(25-60)\%$, the phenomenological spectral density can be roughly
approximated by the "single-pole $+$ continuum states" model
\footnote{ The Borel window is rather small, that may impair the
predicative ability. }. Taking into account all the uncertainties,
we obtain the values of the mass and decay constant from Eqs.(11-13)
\begin{eqnarray}
m_Y&=&(2.46\pm0.16)GeV \, , \nonumber\\
f_Y&=&(8.06\pm 0.87)10^{-4}GeV \, .
\end{eqnarray}

\begin{table}
\begin{center}
\begin{tabular}{c|c|c|c}
\hline\hline
      $M^2(GeV^2)$ &$ \mbox{Pole}(\%)$& $\sqrt{s_0}(GeV)$ &$m_Y(GeV)$\\ \hline
      $1.5-2.0$& $13-28$&$2.3$ & $1.98-2.02$\\      \hline
     $1.5-2.0$& $16-34 $&$2.4 $&$2.05-2.09$ \\     \hline
    $1.5-2.0$& $21-40$ &$2.5$ & $2.12-2.17$\\ \hline\hline
$1.5-2.0$& $25-47$& 2.6 &$2.18-2.24$\\ \hline
          $1.5-2.0$& $30-53$ & $2.7$ &$2.25-2.32$\\ \hline
  $1.5-2.0$& $36-60$ & $2.8$ &$2.30-2.39$ \\ \hline\hline
    $1.5-2.0$&$41-66$ & $2.9$ &$2.36-2.45$\\ \hline
    $2.5-3.0$& $8-14$ & $2.6$& $2.28-2.30$\\ \hline
    $2.5-3.0$& $10-17$ & $2.7$& $2.35-2.38$\\ \hline
     $2.5-3.0$& $12-21$ & $2.8$& $2.43-2.46$\\ \hline
    $2.5-3.0$& $15-25$ & $2.9$ &$2.50-2.53$\\ \hline \hline
\end{tabular}
\end{center}
\caption{ The relation among the Borel parameter,  pole term (or
ground state) contribution, threshold parameter and mass   with the
spectral density  approximated by the perturbative term. }
\end{table}

\section{Conclusion}
In this article, we take the point of view that   $Y(2175)$ be a
 tetraquark state which consists of    color octet constituents, and calculate its mass and decay constant
  within the framework of the QCD sum rules approach. We release standard criterion
  in the QCD sum rules approach and take more phenomenological analysis,  the  value of
  the mass of   $Y(2175)$ is consistent with experimental data; there may be
some tetraquark  components in  the  state $Y(2175)$. On the other
hand, if we retain  standard criterion and take a rather small Borel
window, larger mass than the experimental data can be obtained, the
current $J_\mu(x)$ can interpolate a tetraquark state with larger
mass, or $Y(2175)$ has some components with larger mass. More
experimental data are needed to select the ideal sum rule for the
tetraquark states.  We can take color octet operators as basic
constituents in constructing the tetraquark currents, because the
perturbative term may have dominant contribution,  in other words,
the tetraquark states may consist of color octet constituents rather
than  diquark pairs.

\section*{Acknowledgments}
This  work is supported by National Natural Science Foundation,
Grant Number 10405009,  and Key Program Foundation of NCEPU.

\end{document}